# Joint Characterization of the Cryospheric Spectral Feature Space

Christopher Small
*Lamont Doherty Earth Observatory*
*Columbia University*
*Palisades, NY 10964 USA*
[csmall@columbia.edu](csmall@columbia.edu)

Daniel Sousa
*Department of Geography*
*San Diego State University*
*San Diego, CA 92182 USA*
[dan.sousa@sdsu.edu](dan.sousa@sdsu.edu)

**Abstract**
Spectral feature spaces may be thought of as coordinate systems. When independent spectral dimensions and their corresponding basis vectors reflect salient spectral features, these coordinate systems can be used to characterize a diversity of spectral properties in a physically meaningful way. Multispectral and hyperspectral feature spaces are useful for a variety of remote sensing applications ranging from spectral mixture modeling to discrete thematic classification. In many of these applications, models are used to project the higher dimensional continuum of reflectances (or radiances) onto lower dimensional mappings of the image target's physical properties or categorical composition. In such cases, characterization of the feature space dimensionality, geometry and topology can provide fundamental guidance for effective model design. Utility of this characterization, however, hinges on identification of appropriate basis vectors for the feature space. The objective of this study is to compare and contrast two fundamentally different approaches for identifying feature space basis vectors via dimensionality reduction. In so doing, we illustrate how these two approaches can be combined to render a joint characterization that reveals spectral properties not apparent using either approach alone. We use a diverse collection of AVIRIS-NG reflectance spectra of ice and snow to illustrate the utility of the joint characterization to facilitate both modeling and classification of snow and ice reflectance. Joint characterization is also shown to assist with interpretation of physical properties inferred from the spectra. Spectral feature spaces combining principal components (PCs) and t-distributed Stochastic Neighbor Embeddings (t-SNEs) provide both physically interpretable dimensions representing the global structure of cryospheric reflectance properties as well as local manifold structures revealing clustering not resolved within the global continuum. The joint characterization reveals distinct continua for snow-firn gradients on different parts of the Greenland Ice Sheet and multiple clusters of ice reflectance properties common to both glacier and sea ice in different locations. The clustering revealed in the t-SNE feature spaces, and extended to the joint characterization, distinguishes subtle differences in spectral curvature specific to different spatial locations within the snow accumulation zone, as well as BRDF effects related to view geometry. The ability of the PC+t-SNE joint characterization to produce a physically interpretable spectral feature space revealing global topology while preserving local manifold structures for cryospheric hyperspectra suggests that this type of characterization might be extended to the much higher dimensional hyperspectral feature space of all terrestrial land cover.

**Introduction**

Spectral feature spaces may be thought of as coordinate systems within which high dimensional spectra can be represented on the basis of their most salient features. Analogous to the three dimensional (3D) color spaces in which the continuum of visible colors can be represented as either Cartesian triplets of orthogonal primary color components (e.g. [Red, Green, Blue] = RGB) or cylindrical triplets of orthogonal colorimetric characteristics (e.g. [Hue, Saturation, Value] = HSV), higher dimensional feature spaces can represent a much more extensive continuum of reflectances (or radiances) of non-perceptual colors that can be measured by multispectral and hyperspectral sensors. The utility of multispectral and hyperspectral feature spaces spans a variety of remote sensing applications ranging from spectral mixture modeling to discrete thematic classification. In many of these applications characterization of the feature space dimensionality, geometry and topology can inform the design of models used to project the higher dimensional continuum of reflectances (or radiances) onto lower dimensional mappings of the image target's physical properties or categorical composition.

The tractability of representing a higher dimensional feature space with a lower dimensional model or abstraction is dependent on both the true spectral dimensionality of the continuum of target reflectances and the apparent spectral dimensionality of its representation when sampled in a finite number of discrete spectral intervals. Although rarely considered explicitly in the design of such models for multispectral imagery, the question is central to models for hyperspectral imagery as the transition from undersampling to oversampling narrow absorption features radically changes both the challenges and returns of the modeling process. In studies of hyperspectra, the most commonly used approaches rely on some form of dimensionality reduction (DR) to quantify the dimensionality, and sometimes topology, of the spectral feature spaces being modeled *(Aguila, Efremenko, and Trautmann 2019; Harsanyi and Chang 1994; Khodr and Younes 2011)*. Viewed through this lens, DR algorithms can generally be considered attempts to identify useful spectral feature space basis vectors.

The objective of this study is to compare and contrast two fundamentally different approaches to dimensionality reduction (i.e., basis vector identification) of hyperspectral imagery and illustrate how they can be combined to render a joint characterization of the spectral feature space that reveals spectral properties not apparent using either approach alone. Each algorithm is well-established in the DR literature, but to our knowledge the concept of joint characterization remains largely unexplored. We focus our analysis on this novel approach, using a diverse collection of cryospheric reflectance spectra to illustrate its utility to facilitate both modeling and classification of snow and ice reflectance as well as interpretation of physical properties inferred from the spectra. Snow and ice are particularly illustrative targets because their reflectance often changes continuously and nonlinearly along compositional gradients while maintaining a degree of spatial homogeneity rarely observed with other types of land cover.

## Hyperspectral Dimensionality Reduction and Feature Space Characterization

To extend the color space analogy, the (3-D) Cartesian RGB color space can be generalized to a cylindrical HSV color space within which each distinct (1-D) hue could be considered a spectral EM varying in both (2-D) saturation and value. Higher dimensional spectral feature spaces can similarly be considered vector spaces within which each spectrally distinguishable endmember (EM) represents a measurably distinct dimension along which the amplitude of the EM reflectance increases with distance from a common origin representing a completely dark (absorptive or non-illuminated) target. Along each distinct dimension, the amplitude of the reflectance spectrum is modulated by the illumination and view geometry of the EM material. Spectral mixtures of EMs within the Ground Instantaneous Field of View (GIFOV) of the sensor therefore occupy the hypervolume bounded by the set of independent EM dimensions (assuming linear spectral mixing). In this space bounded by spectrally distinct reflectances, the EMs are effectively basis vectors spanning the space. The apparent dimensionality of the space can be less than the true dimensionality if some compositionally distinct EM materials are indistinguishable from combinations of others when spectrally discretized by the sensor. The apparent (linear; PC) dimensionality of the space can also be more than the true dimensionality if nonlinear mixing is common.

The question of spectral dimensionality and information content is central to the utility of hyperspectral imagery. *(Price 1997)* used a spectral band selection approach to estimate the number of spectral bands necessary to represent the spectral diversity of a set of 45 datasets (13,500,000 spectra) collected by the Airborne Visible Infrared Imaging Spectrometer (AVIRIS) and concluded that 30 to 40 spectral bands should be sufficient to represent nearly all of the information content of the AVIRIS dataset – given the signal/noise characteristics of the instrument circa 1992. However, in an analysis of 1,140,000,000 high altitude (20 m resolution) AVIRIS spectra collected in 1999, *(Boardman and Green 2000)* compared the variance distribution derived from the eigenstructure relative to the instrument's noise floor (circa 1999) to estimate that AVIRIS' spectral dimensionality regularly exceeded 40 and occasionally reached 100. Using a similar approach based on the eigenstructure of the Maximum Noise Fraction (MNF) transform of a variety of low altitude AVIRIS datasets *(Green and Boardman 2000)* estimated dimensionality ranging from 15 for a coastal ocean scene to 57 for an urban scene. More recently, *(Thompson et al. 2017)* compared the PC-derived eigenstructure of 2,500,000,000 cross-track smoothed and unsmoothed high altitude AVIRIS spectra collected over a variety of land cover in California circa 2013 to 2015 to estimate dimensionality ranging from the low 20s to the high 40s for most land cover types, with an average of ~50 for the entire dataset. Using a non-parametric approach based on random matrix theory applied to a 1997 high altitude AVIRIS dataset from a spectrally diverse hydrothermal complex in Cuprite NV, *(Cawse-Nicholson et al. 2013)* estimated dimensionality ranging from 22 to 30; encompassing the range of previous estimates for the same complex.

Global analysis of spectrally diverse collections of broadband spectra reveal a consistent feature space topology in which many landscapes can be represented as mosaics of rock and soil substrate, green and nonphotosynthetic vegetation and dark/absorptive features like water and shadow *(Small 2004a)*. Evaporites, ice/snow and shallow marine substrates (e.g. reefs) tend to form distinct (frequently non-linear) subspaces but can also be represented as continua spanned by compositionally distinct EMs modulated by BRDF effects radiating away from a common dark origin. For a wide variety of landscapes a simple Substrate, Vegetation, Dark (SVD) spectral mixture model of broadband reflectance was found sufficient to describe 97% of 100,000,000 Landsat spectra with misfit < 0.05 *(Small and Milesi 2013)*. In part, this is a consequence of the near-orthogonality of the S, V and D spectral continua. However, comparative analysis of coincident Landsat multispectral and AVIRIS hyperspectral observations in a diverse collection of agricultural, grassland, forest, and savanna settings indicate that hyperspectral feature space topology is also dominated by the spectral continua of substrates and vegetation with amplitude modulated by BRDF effects – at least for low-order dimensions as determined by a linear variance-based decomposition *(Sousa and Small 2018)*.

Even though imaging spectrometers like AVIRIS can resolve narrow band absorptions, the covariance structure of the hyperspectral feature space is dominated by the spectral continuum more so than the lesser contributions of individual absorption features. This illustrates the aforementioned bias that results from using dimensionality reduction criteria that treat "global" variance (defined over the entire feature space) as a proxy for information content (e.g. Principal Components). While the shape of the spectral continuum captures important distinctions among broad categories of spectra (like rock and soil substrates, green vegetation, ice and snow), more subtle distinctions related to diagnostic narrowband absorption features within each category generally do not contribute enough variance to influence the topology of the low order dimensions of the PC-projected feature space. Much of the potentially useful information content of hyperspectral feature spaces is thus effectively suppressed by DR methods that prioritize high variance global structure (e.g. spectral continuum) at the expense of any lower variance local features (e.g. narrowband absorptions) that may be manifest in the form of nonlinear manifolds embedded within the higher dimensional feature space.

The realization that such local (in feature space) scale manifolds may be important for certain applications has led to the development of dimensionality reduction algorithms designed to identify them. Such algorithms are commonly given the moniker "manifold learning", alluding to the possible existence of lower dimensional manifolds embedded within a higher dimensional feature space. Examples include Isomap (*(Tenenbaum et al., 2000)*; applied to hyperspectral imagery by *(Bachmann et al., 2005)*, locally linear embedding *(Roweis and Saul, 2000)*, Laplacian eigenmaps *(Belkin and Niyogi, 2003)*, Hessian eigenmaps *(Donoho and Grimes, 2003)*, and diffusion maps *(Coifman and Lafon, 2006)*. For an introductory overview of manifold learning, see *(Izenman, 2012)*. While each of these algorithms operates using different quantitative metrics and mathematical constructs, the overall conceptual framework of each is the same: an attempt to provide an efficient low-dimensional representation of high-dimensional local manifolds. Most of these algorithms operate nonlinearly, and can thus require considerable computational resources. The objective of joint characterization of the global feature space structure and local manifolds within the feature space is to more accurately represent the diversity of information content resolved by the full set of spectral measurements.

Among non-linear manifold learning approaches, the t-distributed Stochastic Neighbor Embedding (t-SNE; *(van der Maaten and Hinton, 2008)*) algorithm has gained popularity in recent years due to its ease of use and quality of visualization. Briefly, t-SNE attempts to find the optimally faithful low-dimensional (usually 2-D) representation of a higher dimensional dataset. Euclidean distances between all pairs of data points are converted into conditional probabilities, and faithfulness between the low-D and high-D representations is quantified using the Kullback-Leibler divergence. The t-distributed aspect of t-SNE was designed to avoid some of the pitfalls of the earlier Stochastic Neighbor Embedding (SNE; *(Hinton and Roweis, 2002)*), specifically a difficult cost function and the "crowding problem", a challenge related to the curse of dimensionality. For the purposes of this work, t-SNE was chosen as the exemplary manifold learning technique due to its popularity, ease of implementation, and encouraging initial results.

Although it is clear that both global and local algorithms can be useful for analysis of high dimensional data, implementation is often viewed as a choice: either explicitly attempt to preserve "local" manifold structure or focus on high variance global features like spectral continuum shape. Instead of accepting this "either/or" dichotomy, we instead choose to consider an "and" approach: joint characterization using both global and local metrics. Specifically, we compare, contrast and combine principal component (global) and t-SNE (local) feature spaces for a composite of several hyperspectral datasets spanning a range of cryospheric environments. As introduced by *(Sousa & Small 2021)*, this joint approach can leverage the benefits of both local and global algorithms – in principle.

**Data**

In this analysis, we use a set of AVIRIS-NG lines collected on and around the Greenland Ice Sheet to construct a composite hyperspectral feature space intended to represent a diversity of cryospheric spectra. The composite space consists of three peripheral ice sheet transects spanning snow-firn-ice continua on distinct gradients between accumulation and ablation zones, and two near-coastal lines sampling sea ice with varying amounts of snow cover and open water. In addition, we include a single AVIRIS-NG line from the Indian Himalaya spanning a variety of snow-covered mountain glaciers separated by topographic ridges with a wide range of slopes and azimuths. This line contributes a wider range of view and illumination geometries than the Greenland lines, in addition to a greater amount of wind blown dust than is typically found on most parts of the Greenland Ice Sheet.

The Airborne Visible / Infrared Imaging Spectrometer – Next Generation (AVIRIS-NG) is a state-of-the art instrument built and maintained by NASA's Jet Propulsion Lab (JPL). AVIRIS-NG data provide 5 nm sampling over the range 380 to 2510 nm, with 600 cross-track elements. Ground sampling distance varies with ground elevation and aircraft altitude but generally ranges from 0.3 m to 8.0 m for low altitude collections. AVIRIS-NG data are available from JPL's web portal at: https://avirisng.jpl.nasa.gov/dataportal/. All AVIRIS-NG lines used in this analysis were atmospherically corrected and calibrated to surface reflectance by the AVIRIS team as provided by the portal.

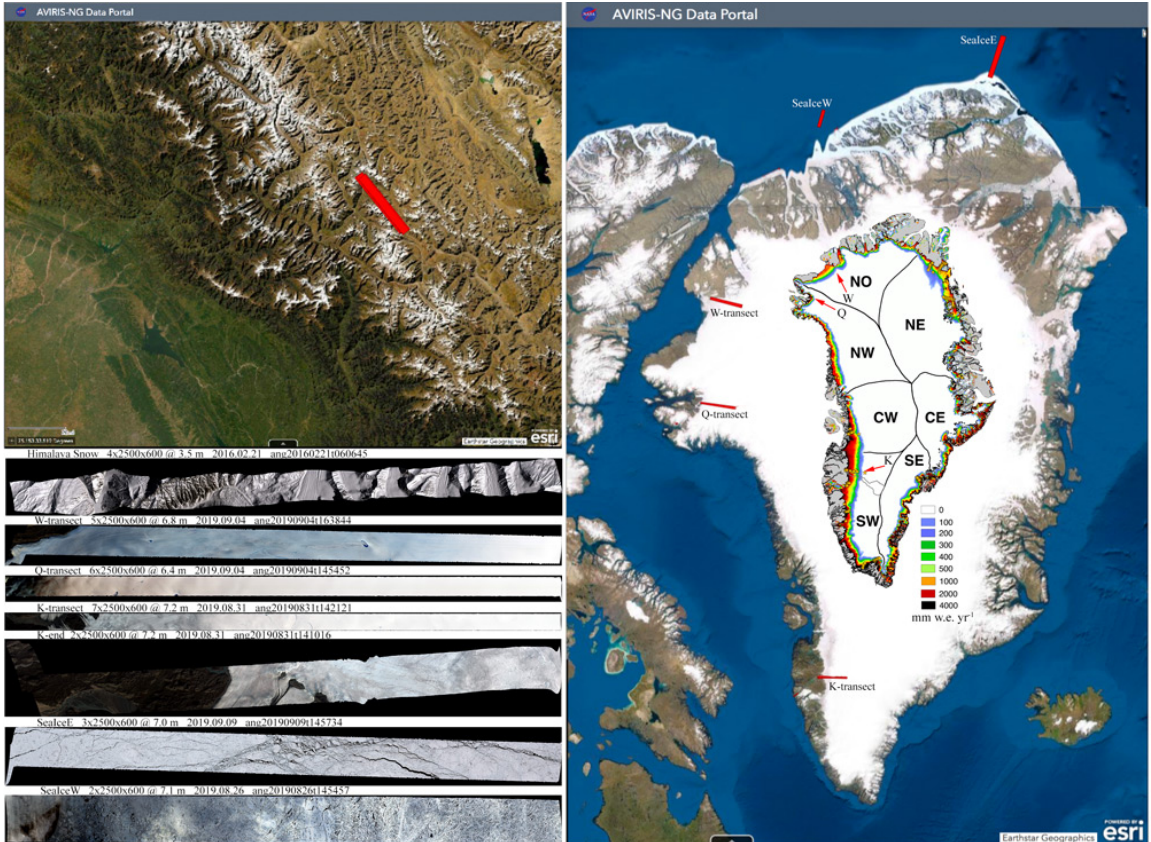

*Figure 1 Index maps with natural color composites of AVIRIS-NG acquisitions used to construct the cryospheric spectral cube and feature space. All three Greenland transect lines span transitions from ablation zones at western ends to accumulation zones at eastern ends. Inset modeled runoff map from Noël et al. (2019) shows each transect spanning distinct snow-ice gradients. The SeaIce E & W lines both contain pack ice spanning a range of fragmentation, open water and snow cover conditions. The Himalayan snow line contains uniform snow cover on a variety of mountain glaciers and rock/soil substrate ridges having a wide range of slopes and aspects giving rise to a wide range of view and illumination geometries. Only cloud-free fully illuminated sections of lines are included in the composite mixing space.*

The three ice sheet transect lines were chosen to span snow-firn-ice gradients of varying width (Figure 1) and varying local climatic conditions. The Kangerlussuaq (K) and Qaanaaq (Q) transect lines image snow and firn over most of their length, with exposed ice only at the westernmost ends, whereas the Washington (W) transect images ice over most of its length with snow and firn only at its southeasternmost end. Cloud free conditions allowed the full length of each transect line to be used. The sea ice lines (W and E) image pack ice with interspersed fractures exposing open water. Partial cloud cover of the western line reduced the usable length of cloud free coverage. The Himalaya (H) snow line images pervasive snow cover with some areas of exposed rock and soil substrate on steeper slopes.

Spatial subsets of each AVIRIS-NG line were selected to avoid data gaps and cloud cover. For each line, a sequence of 2500 x 600 pixel spatial subsets was selected to cover as much of the line as was usable. The number of subsets, given in Figure 1, was determined by the length of each line. These subsets are arranged in sequentially in order of line direction (West to East for the transect lines) to form a rectangular hyperspectral cube progressing from the sea ice lines to the transect lines (South to North) to the Himalaya line (Figure 1). The pixel resolution of the Greenland lines ranges from 6.4 to 7.2 m while the Himalaya line has a resolution of 3.5 m.

**Analysis**

The analysis begins with a comparison of a global spectral feature space derived from the three low order principal components (PCs) and a local feature space derived from a 2D t-SNE projection of the full cryospheric composite. Unlike the unique and invertible PC transform, the t-SNE projection algorithm contains a stochastic seeding that produces a reduction of the local manifold embedding of the higher dimensional feature space. For this reason, we refer to the 2D output of any single t-SNE reduction as a realization to reflect its nonuniqueness. In order to identify the local manifold structure of the feature space in a more robust and repeatable manner, we generate several t-SNE realizations of the AVIRIS-NG composite and compare the low order PCs of this collection with the PCs of the original composite. The question of convergence of the PC($[t\text{-}SNE_1 \ldots t\text{-}SNE_N]$) for varying sets of N realizations is discussed in the Appendix.

The joint characterization of the composite feature space employs the strategy given by (Sousa and Small 2021) to place the local scale manifolds identified by t-SNE in the context of the global scale topology given by the low order PCs. This approach retains the physically interpretable context of the global feature space topology while preserving the more subtle local scale manifold structure(s) captured by t-SNE. We combine the local manifold embeddings identified by multiple 2D t-SNE realizations with the global structure identified by the low order PCs of the composite by rendering a series of 2D projections combining a low order PC dimension with a PC(t-SNE) dimension. This allows the continua derived from the PC projection to be combined with the manifold clustering identified by PC(t-SNE) to show both the continuous and discontinuous structures within the higher dimensional feature space. Spectral characteristics of both continua and clusters within the joint feature space are illustrated by comparison of individual and cluster mean spectra.

Extending the joint characterization from the composite cryospheric feature space to the distinct feature spaces of individual AVIRIS-NG lines makes it possible to compare and contrast feature space topologies of the three transect lines spanning snow-firn-ice gradients on the western edge of the Greenland Ice Sheet. This, in turn, facilitates comparison of the common spectral features that control the topology of the composite feature space given by the low order PCs with potentially different spectral features that control the topologies of individual lines' (therefore geographies) feature spaces. In both the composite feature space and the individual lines' feature spaces, the PCs provide a more physically interpretable structure, generally controlled by the distribution of spectral continua determining the covariance structure, while the low order PC(t-SNE) dimensions would be expected to identify the same local manifolds in the composite space and the individual lines' feature spaces.

**Results**

Comparing the PC-derived global structure with the t-SNE-derived local structure for the AVIRIS-NG composite illustrates the stark contrast between the DR approaches (Figure 2). The 3D PC feature space, shown as orthogonal 2D projections of the 3 low order PCs, is continuous throughout, with only the Himalaya snow mixing continuum appearing distinct from a single ternary continuum containing all the other lines. The ternary structure is bounded by spectral endmembers (EMs) at the apexes corresponding to fine snow, ice and shadow/water. With the exception of two clusters near the snow apex and a cluster midway between the snow and ice apexes, the internal structure of the ternary part of the feature space is topologically continuous with amplitudes and shapes of spectra varying continuously throughout the space as illustrated in Figure 2. The change of spectral amplitude and curvature along the continuum from fine snow to ice shows the well known reduction in Near IR reflectance and broadband albedo predicted from theory *(Bohren and Barkstrom 1974; Wiscombe and Warren 1980)* and observed in field studies (e.g. *(Dadic et al. 2013)*). Linear mixing lines extend from both ice and snow apexes to the dark apex representing both deep water (on the sea ice lines) and deep shadow. Shadow on the Greenland transect lines is primarily associated with glacier crevasse while the shadow on the Himalaya line results from steep topography relative to the solar geometry at the time the line was collected. The distinct mixing continuum of the Himalaya snow line is a result of variations in flux density arising from variations in incidence angle on a wide range of glacial valley slopes, as well as subpixel mixing of snow and rock/soil substrate in areas with incomplete snow cover. Short wavelength visible absorption is also conspicuous on the Himalaya line, which we attribute to light absorbing impurities such as mineral dust and carbon *(Warren 2019; Warren and Wiscombe 1980)*. Because black carbon can reduce the visible albedo of snow, we cannot attribute the mixing lines with the dark endmember to illumination alone, but because of the multiplicity of other factors that can also reduce snow reflectance (e.g. *(Warren 2013) (Painter et al. 2013)*), we refrain from speculation as to effects other than the known illumination geometry from surface topography (e.g. crevasse) observable on the full resolution imagery. We base our interpretation of the spectra in the feature space on the criteria and characteristics of cryospheric spectra discussed by *(Dozier et al. 2007; Dozier and Painter 2004) (Bohn et al. 2021)* and *(Warren 2019)*.

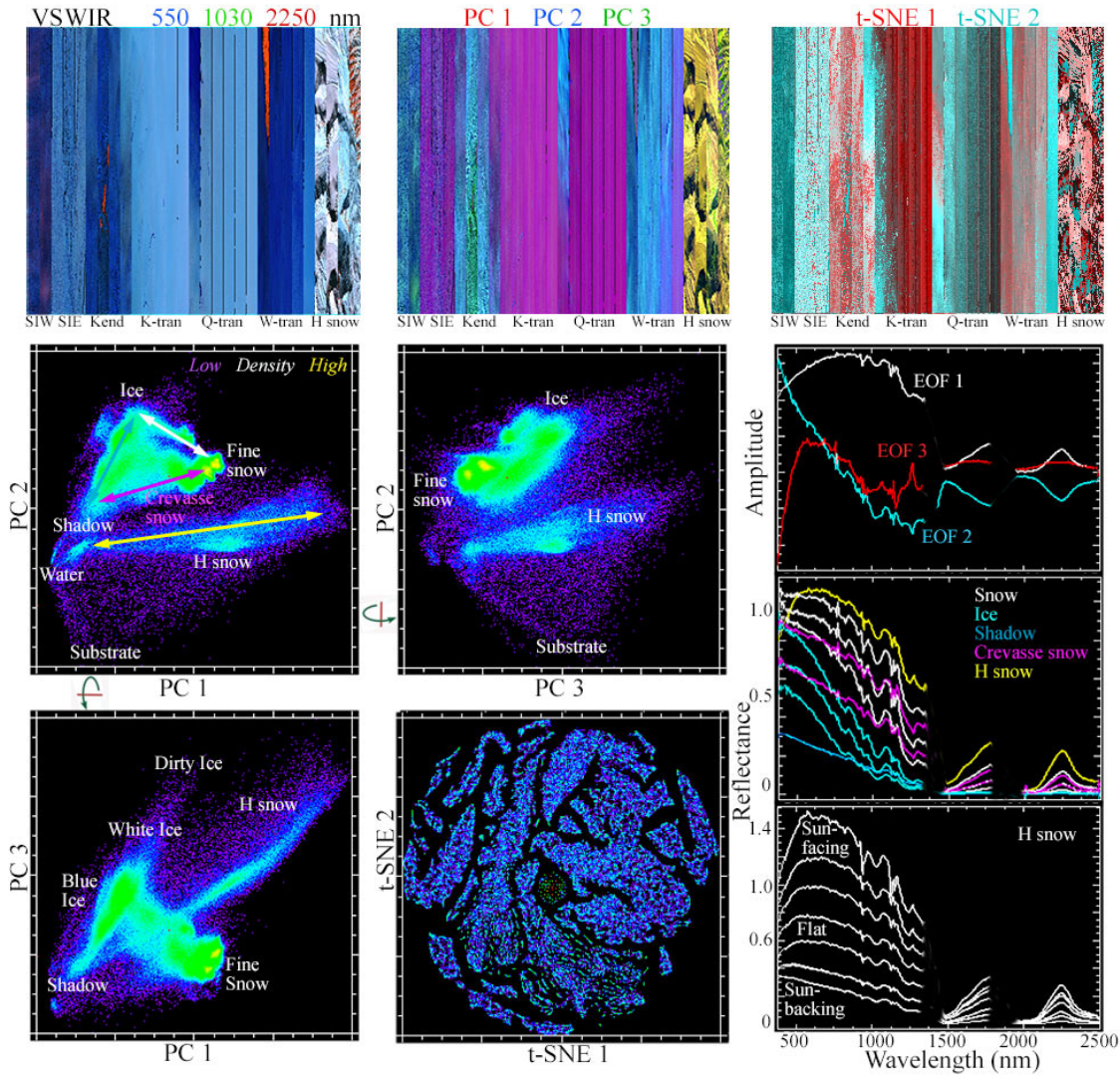

*Figure 2 Cryospheric spectral feature spaces rendered from low order PC and t-SNE projections of the AVIRIS-NG composite. The PC space is continuous and bounded by compositionally distinct spectral endmembers (EMs). The first 3 dimensions of the PC space account for 94.2%, 4.4% and 0.65% of total variance, making the space effectively 2 dimensional. The PC space topology represents the linear combinations of the corresponding Empirical Orthogonal Functions (EOFs) that produce the continuum of spectra in the feature space. The similarity of EOFs 1 & 2 to snow and ice spectra gives rise to the continua of spectral shape and amplitude along binary gradients bounding the triangular upper part of the feature space. This upper space spans gradients of clean snow and ice with amplitude modulated by shadow, water and view/illumination geometry. The uniform dirty snow cover in the H snow line forms a separate continuum between fully shadowed slopes and specular reflections (R > 1). Small areas of exposed rock/soil substrates form a mixing continuum between a substrate EM and the H snow mixing trend. In contrast to the continuous PC space, the clustered t-SNE space preserves fine scale local structure of the higher dimensional feature space but gives no indication of clusters' physical properties. Shapes and positions of clusters in 2D t-SNE space are effectively random but the segregation of the clusters suggests the presence of*

*distinct fine scale spectral features that contribute very little (< 1% total) variance, dispersed among the many higher order dimensions of the PC space.*

In contrast to the continuous PC space, the 2D t-SNE feature space is strongly clustered with ~10 large distinct clusters interspersed among several much smaller clusters (Figure 2). Unlike the physically interpretable topology of the PC-derived global space, the positions, shapes and sizes of the clusters within the t-SNE space bears no obvious relationship to their spectral properties.

The stochastic seeding of the t-SNE algorithm naturally raises questions about the uniqueness and repeatability of the clusters it identifies. We address this issue by generating multiple realizations of the 2D t-SNE feature space and compute the Principal Components of a sequence of realizations to determine whether cluster membership is consistent across realizations. Figure 3 (column 1) shows 5 realizations of 2D t-SNE feature spaces for comparison. It is immediately apparent that the number and size distribution of clusters is similar for all the realizations. In some cases larger clusters appear to have similar shapes over multiple realizations. Most obvious is the sparse circular cluster easily identified in all the 2D t-SNE spaces. This cluster corresponds to null spectra areas where the imaged swath was slightly less than 600 pixels. Comparing 2D PC spaces for multiple t-SNE realizations (PC(t-SNE)) shows a discontinuous feature space containing multiple elongate continua combined with several distinct clusters (Figure 3, column 2). Increasing the number of realizations included in the PC increases both the degree of clustering and the continuity of the clusters. The sparse circular cluster of null spectra seen in the 2D t-SNE spaces coalesces into a denser spherical cluster in the PC(t-SNE) space. The low order PCs of a sequence of 30 2D t-SNE realizations shows considerably more distinct continua and clusters when compared to the PCs of only 4 t-SNE realizations (Figure 3, column 3), although higher order dimensions are less continuous and clustered than the lower order dimensions. Finally, combining the low order PCs of the AVIRIS-NG composite with the low order PC(t-SNE) dimensions provides the joint characterization we seek (Figure 3, column 4). The spectral amplitude controlled by PC 1 represents the variation in illumination spanning the snow, ice and dark binary mixing continua while PC(t-SNE) 1 retains both the clustering and continuum structures.

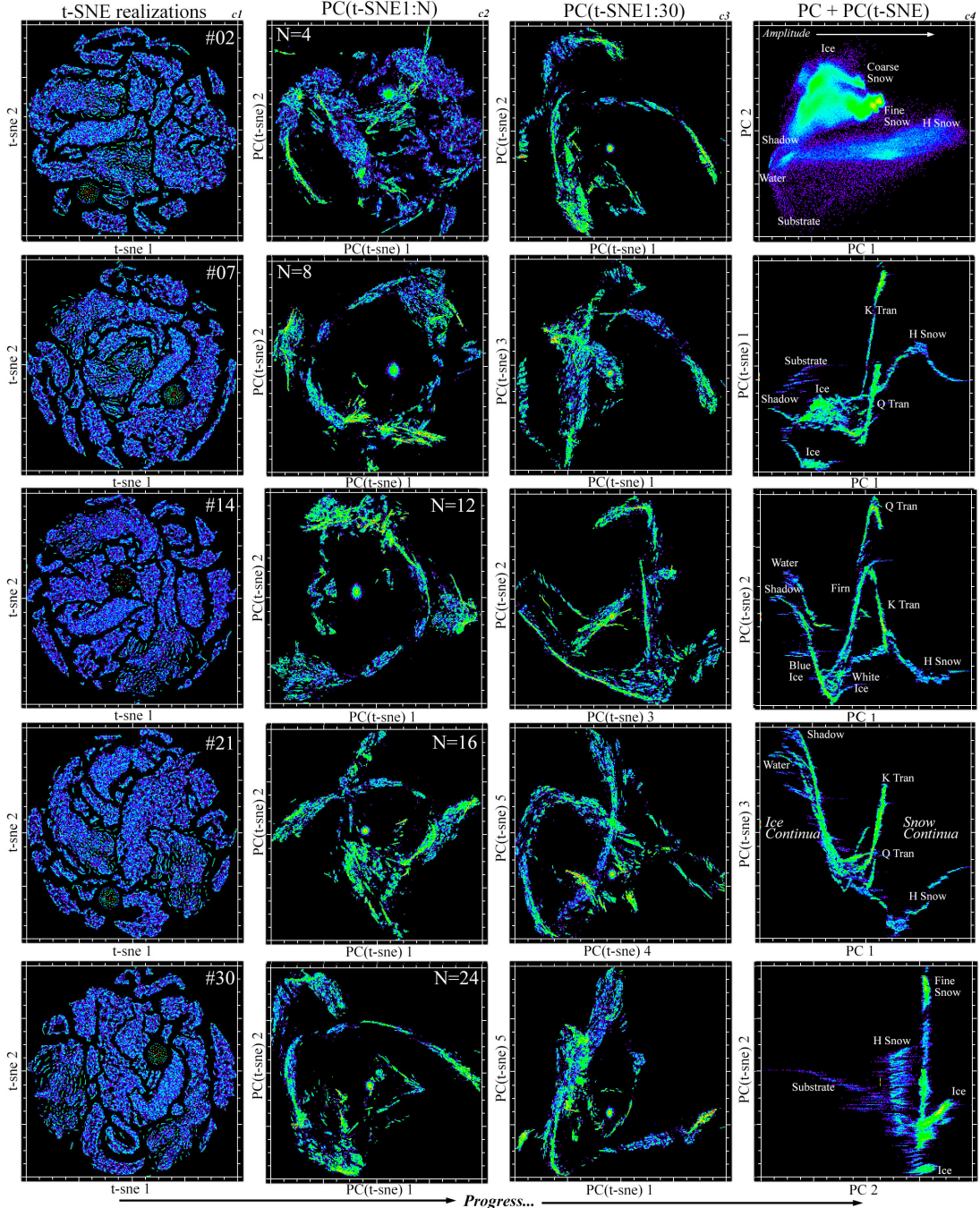

*Figure 3 Multiscale PC+t-SNE spectral feature spaces. Multiple 2D t-SNE realizations (c1) have distinct clustering patterns, but similar size distributions. PCs of increasing numbers of 2D t-SNE realizations (c2) show progressive coalescence into smaller numbers of distinct elongate manifolds. Projections of low order PCs from a set of 30 2D t-SNE realizations (c3) maintain this elongate manifold structure although clusters become more dispersed in higher order PCs. Joint PC+ PC(t-SNE) characterization (c4) combines the global spectral continua of the PC feature space (Fig. 2) with the local clustering of the 2D t-SNE feature spaces accentuated in the low order PCs of 30 t-SNE realizations. The t-SNE manifold continuity persists, but can be interpreted in terms of*

*the global continua of spectral properties reflected in the topology of the PC space. The perplexity of the t-SNE realizations is 30 in all cases.*

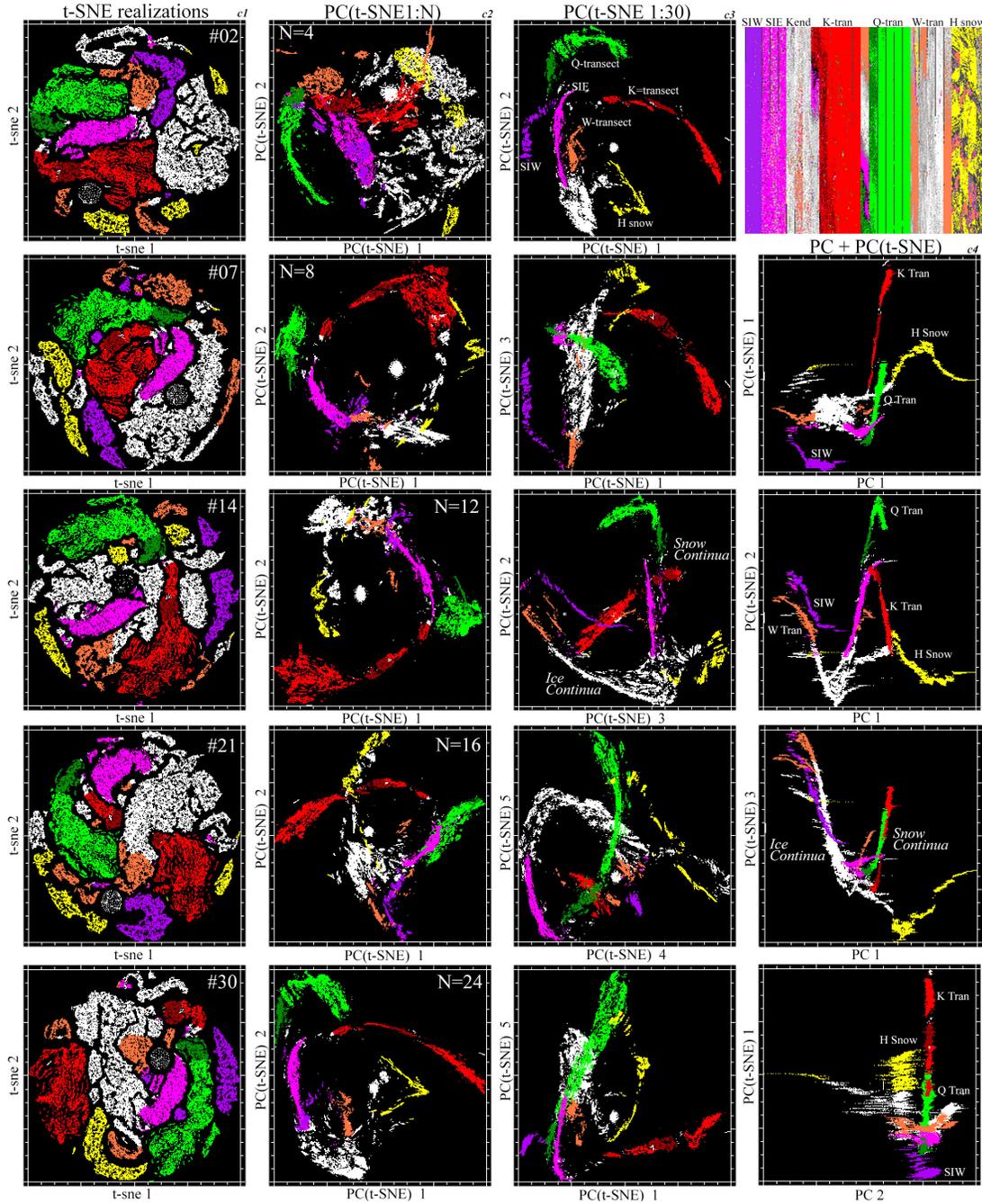

*Figure 4 Spectral feature space dis-continua correspond to geographic compositional characteristics. Adjacency of distinct clusters forming elongate limbs in PC(t-SNE) space (c2 & c3) correspond to distinct clusters in each realization of t-SNE space (c1) and continua of snow and ice reflectance on specific AVIRIS-NG lines In the PC+PC(t-NE) space (c4) the PC dimension (x) preserves continuous variations in global structure related to spectral shape and amplitude while dis-continuity of distinct clusters in the PC(t-SNE) dimension (y) preserves fine scale local features associated with distinct*

*manifolds in the higher dimensional spectral feature space. Note that the geogaphic specificity of some AVIRIS NG lines is preserved in the t-SNE spaces but not in the low order PC space continuum of Fig. 2. The local spectral similarities that yield clusters in t-SNE space do not contribute enough variance to be discernable in the continuous PC space. In contrast, individual ice clusters span multiple AVIRIS NG lines.*

Mapping individual labeled clusters from PC(t-SNE) space back into the original 2D t-SNE feature spaces confirms that distinct clusters in PC(t-SNE) space do indeed correspond to individual clusters in each realization of the 2D t-SNE space (Figure 4). Similarly, mapping the same clusters back into the geographic space of the composite reveals that several (but not all) of the most distinct clusters in the PC(t-SNE) and 2D t-SNE spaces map onto snow covered sections of individual transect lines, as well as the Hsnow line. This geographic consistency provides immediate confirmation that the clusters identified by t-SNE correspond to physical similarity of spectral properties because no explicit geographic information is provided to the t-SNE algorithm. The spectral properties on which t-SNE basis its clustering clearly correspond to physical differences in the spectral properties of snow on the different lines.

A closer examination of the PC+PC(t-SNE) feature space reveals a clear distinction between snow and ice continua in the joint characterization. Whereas the PC+PC(t-SNE) space is composed of both distinct continua and clusters, the distinct continua correspond to snow on individual AVIRIS-NG lines in different geographies while the less distinct clusters corresponding to firn and ice are not line-specific with individual clusters spanning multiple lines (Figure 5). This suggests that the compositional gradients in snow are spectrally distinct from one location to another while the ice continua and clusters include ices on all three transect lines and both sea ice lines.

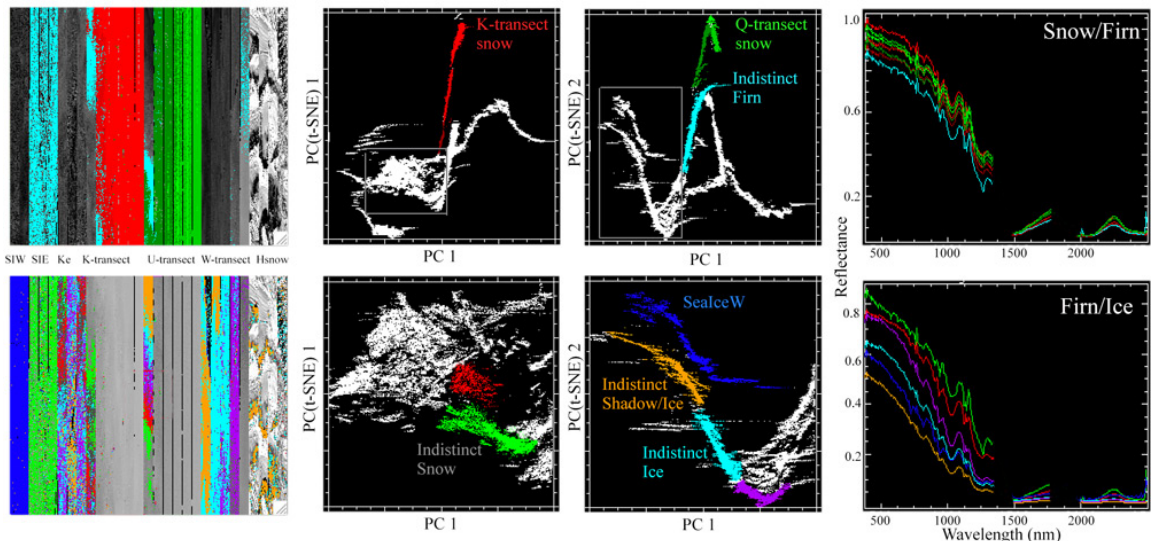

*Figure 5 Continuous distinct snow-firn gradients and discontinuous indistinct firn-ice distributions. In the multiscale PC+PC(t-SNE) feature spaces, snow gradients on the eastern K and Q transects (top) have similar reflectance with very little change in amplitude but form distinct clusters in the PC(t-SNE) dimensions. However, the separate cluster corresponding to indistinct firn (cyan) represents much of SeaIceE and parts of*

*all 3 transects. Clusters from enlarged subspaces (bottom) span a much wider range of lower reflectances but all except SeaIceW (blue) are present in at least 2 locations and each transect contains at least 3 indistinct reflectances.*

The spectral continua of the two snow-dominated transect lines (K and Q) and the H snow line correspond to quantifiable differences in spectral curvature. Most obvious is the distinction between clean and dusty snow on the Greenland transects and Himalaya lines (respectively). The conspicuous short wavelength visible absorption seen on all the Hsnow spectra (Figure 2) is absent on any of the Greenland transect snow spectra. The more subtle differences between the K and Q transect snow spectra are illustrated in Figure 6. While the reflectance of spectra from both lines is comingled at SWIR wavelengths, they begin to segregate in the NIR and are well separated at visible wavelengths. Estimates of snow grain size, derived from the 1030 nm absorption band area relation of *(Nolin and Dozier 2000)*, span similar ranges of westward-coarsening grain radii (450-700 μm) over the length of the snow-firn covered part of both transect lines (Figure 6 inset). It is also known that the liquid water content of snow affects the shape and depth of the 1030 nm absorption feature *((Hyvarinen and Lammasniemi 1987; Nolin and Dozier 1993; Green et al. 2002)*. The range of grain sizes estimated for both K and Q transects are consistent with grain sizes measured for melting snow elsewhere on the Greenland ice sheet by *(Nolin and Dozier 2000)*. The most obvious difference between the K and Q transect snow spectra appears to be the slope and curvature of the spectral continuum between ~500 to 700 nm.

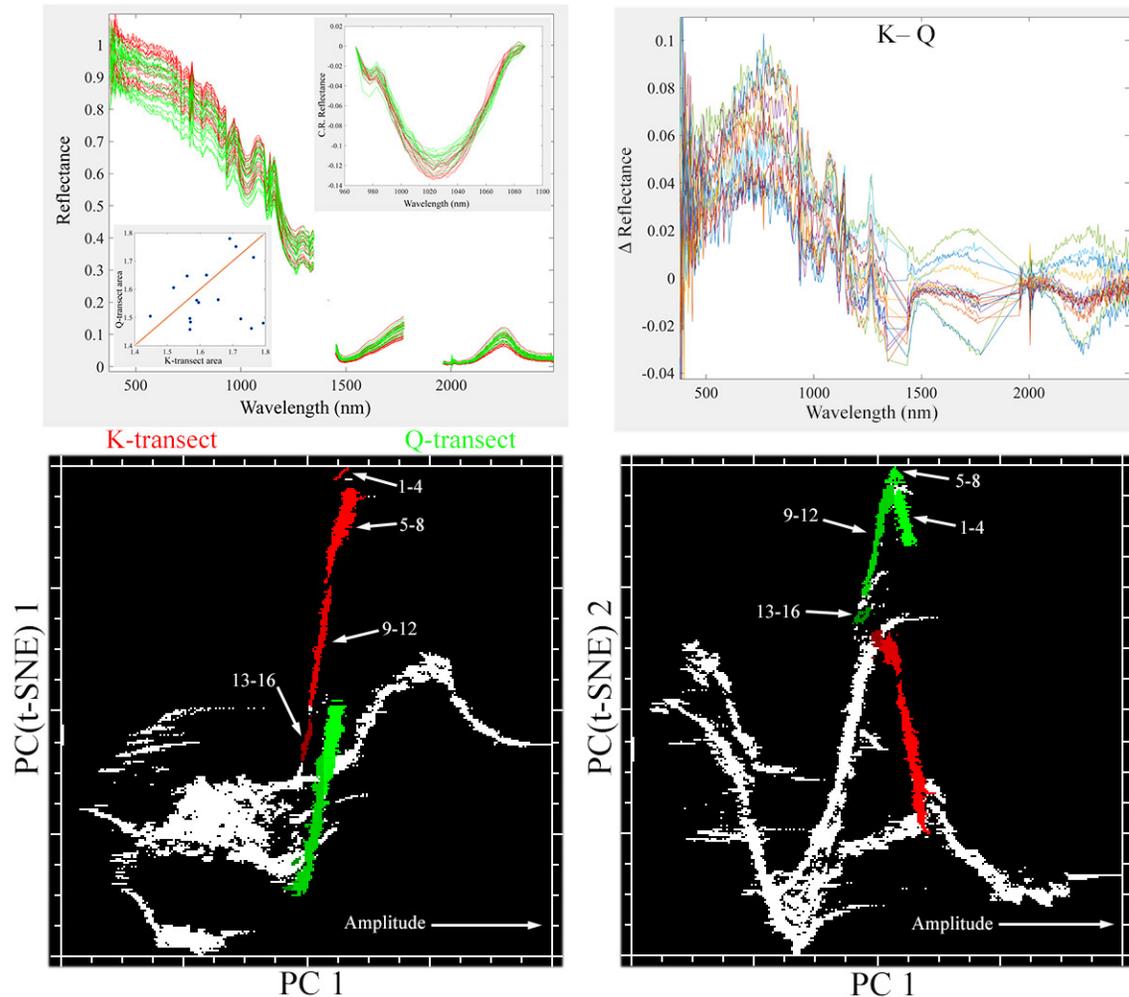

*Figure 6 Comparison of K and Q transect spectral continua. Distinct K and Q snow spectra span similar amplitudes and are fully comingled through the SWIR (>1000 nm) with different curvature in the VNIR. Integrated areas of continuum-removed 1030 nm absorptions (inset) are similar for sequential sets of 16 K & Q spectra, with area corresponding to grain radii ranging from 450 mm to 700 mm. Radii generally increase westward with decreasing spectral amplitude, but the trend is not monotonic.*

Comparison of the PC and PC(t-SNE) feature spaces for individual transect lines reveal additional distinctions not apparent in the composite feature space. Figure 7 shows each transect having a PC feature space containing snow-firn-ice continua composed of two limbs. One limb spans a snow-firn continuum, connecting to a second limb spanning a continuum between blue and white ice. Each line's snow-firn continuum intersects its ice continuum at a different point. In contrast, all three lines' PC(t-SNE) feature spaces contain elongate continua for snow-firn and distinct clusters for ice – consistent with the dichotomy seen in the composite feature space. The snow-firn continua on the K and Q transects both correspond to spectral amplitude and both continua taper to form distinct apexes at the high amplitude end. These distinct apexes correspond to opposite sides of the AVIRIS-NG swath, distinguishing small amplitude differences in the sun-facing and sun-backing reflectances, presumably resulting from BRDF effects resolved by the AVIRIS-NG sensor (e.g. (*Painter and Dozier 2004)*). The PC(t-SNE) continuum on the W transect corresponds to amplitude variation in firn reflectance, but does not distinguish separate apexes for sun-facing and sun-backing reflectances as K and Q do with snow. Combining the PC1 + PC(t-SNE)1 dimensions in the joint characterization projects the snow-firn continua and ice clusters of the PC(t-SNE) manifolds onto the spectral amplitude continuum of the PC space for each line, further highlighting the contrast between the continuous and clustered reflectances.

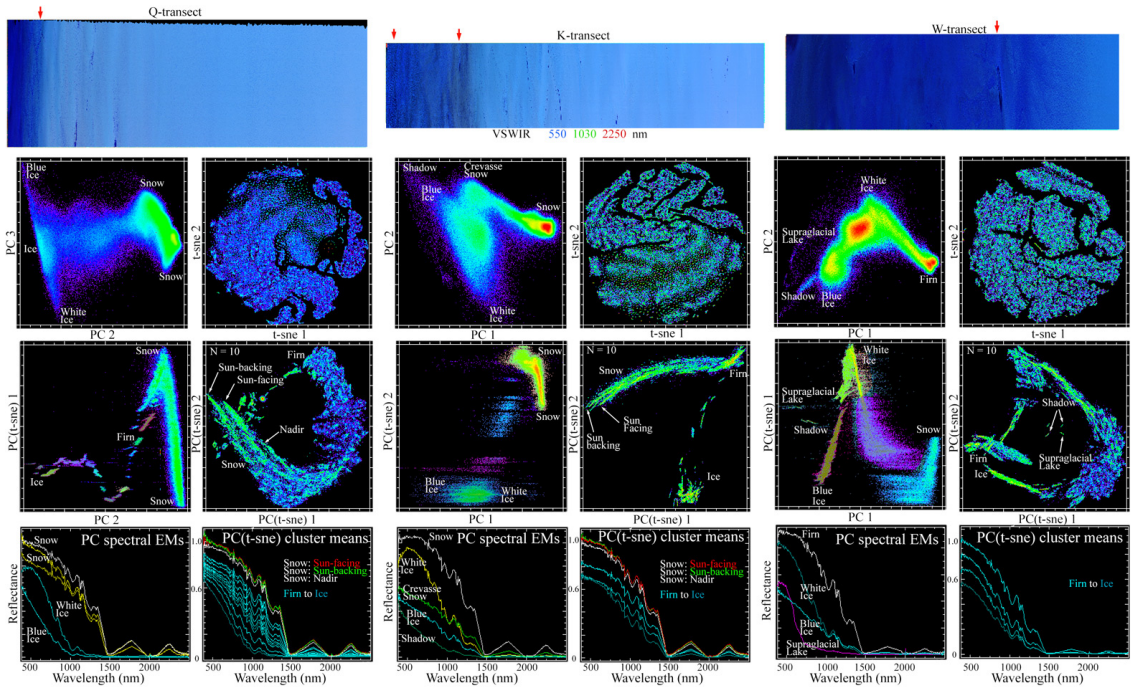

*Figure 7 Multiscale spectral feature spaces for K, Q & W transect lines. All 3 have continuous PC feature spaces with snow-ice continua, however the snow-ice continuum intersects the continuum spanning white and blue ice at different locations for each transect. Note that individual transect lines' continuous PC spaces occupy subspaces of the more continuous full composite space (Fig. 2), In contrast, the t-SNE space of each line is dominated by a single large cluster corresponding to a snow/firn continuum and several smaller clusters corresponding to firn/ice discontinua. Separable clusters are highlighted in the PC+PC(t-SNE) spaces, with cluster mean spectra showing the continuous variation in spectral amplitude and curvature represented by the PC dimension. The snow on the K and Q transect lines both clearly distinguish the sun-facing and sun-backing sides of the swath in the PC(t-SNE) spaces - even though the amplitude differences contribute too little variance to be distinguished in the PC spaces. Compressed length AVIRIS-NG images (top) show locations (arrows) of full resolution examples in Figs. 8-11.*

Full resolution subsets of each AVIRIS-NG line illustrate the diversity of reflectances in the glacier ice of the transect lines (Fig. 8-11). The < 7 m resolution of the lines is sufficient to resolve individual crevasse, and the meter scale spectral diversity of white, blue and dirty glacier ice. The observation that this diversity spans quasi-linear continua in the individual lines' PC spaces suggests that these ice types may be spectrally separable and therefore estimable by inversion of linear spectral mixture models.

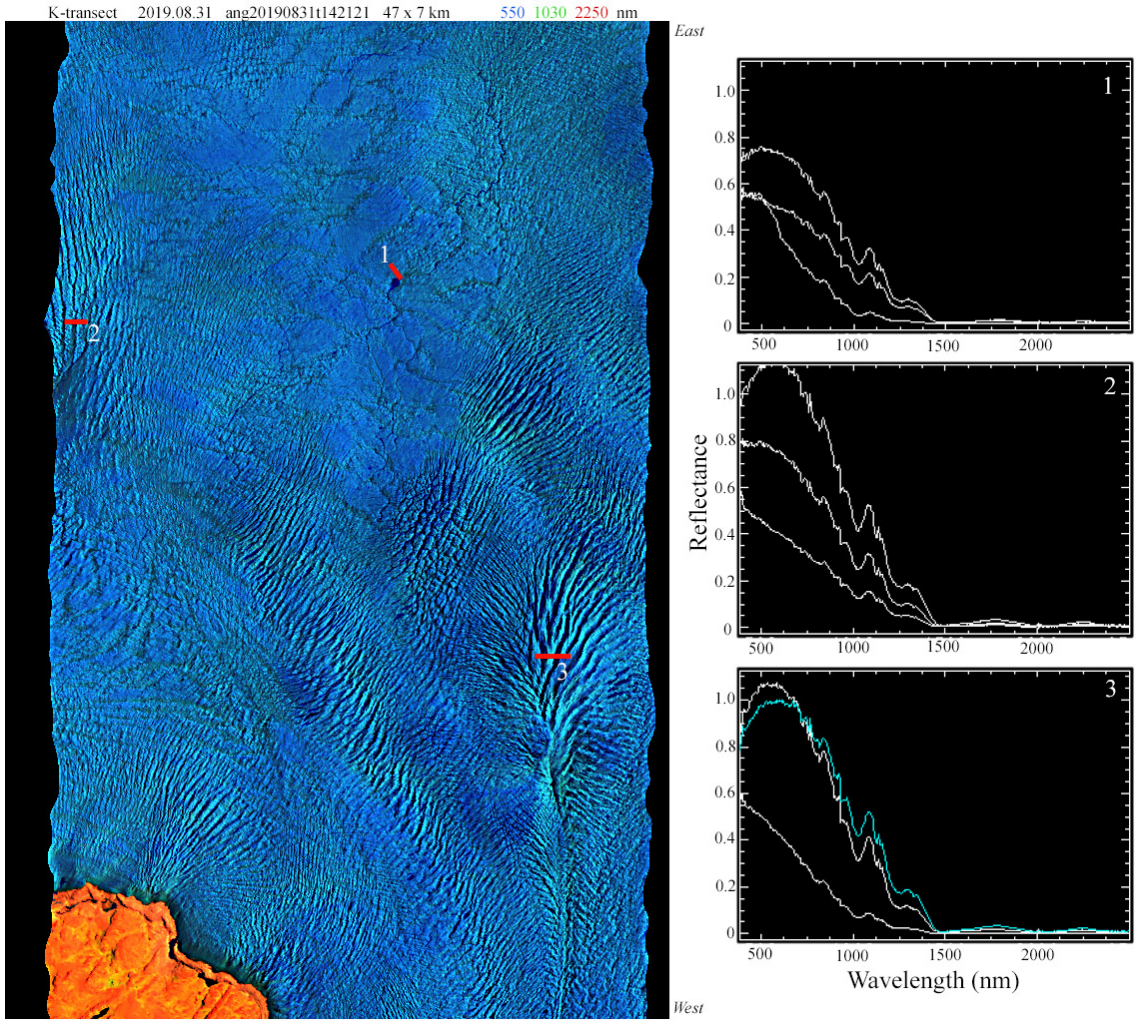

*Figure 8  Full resolution AVIRIS-NG swath and spectra from the western K-transect. Variations in reflectance and illumination geometry reveal crevasse texture and surface slope and aspect.  Example spectra from adjacent, or near-adjacent, pixels show meter to decameter scale variations in ice reflectance that would contribute to the spectral mixing occurring within decameter resolution imagery.  Ground sample distance is 7.2 m.*

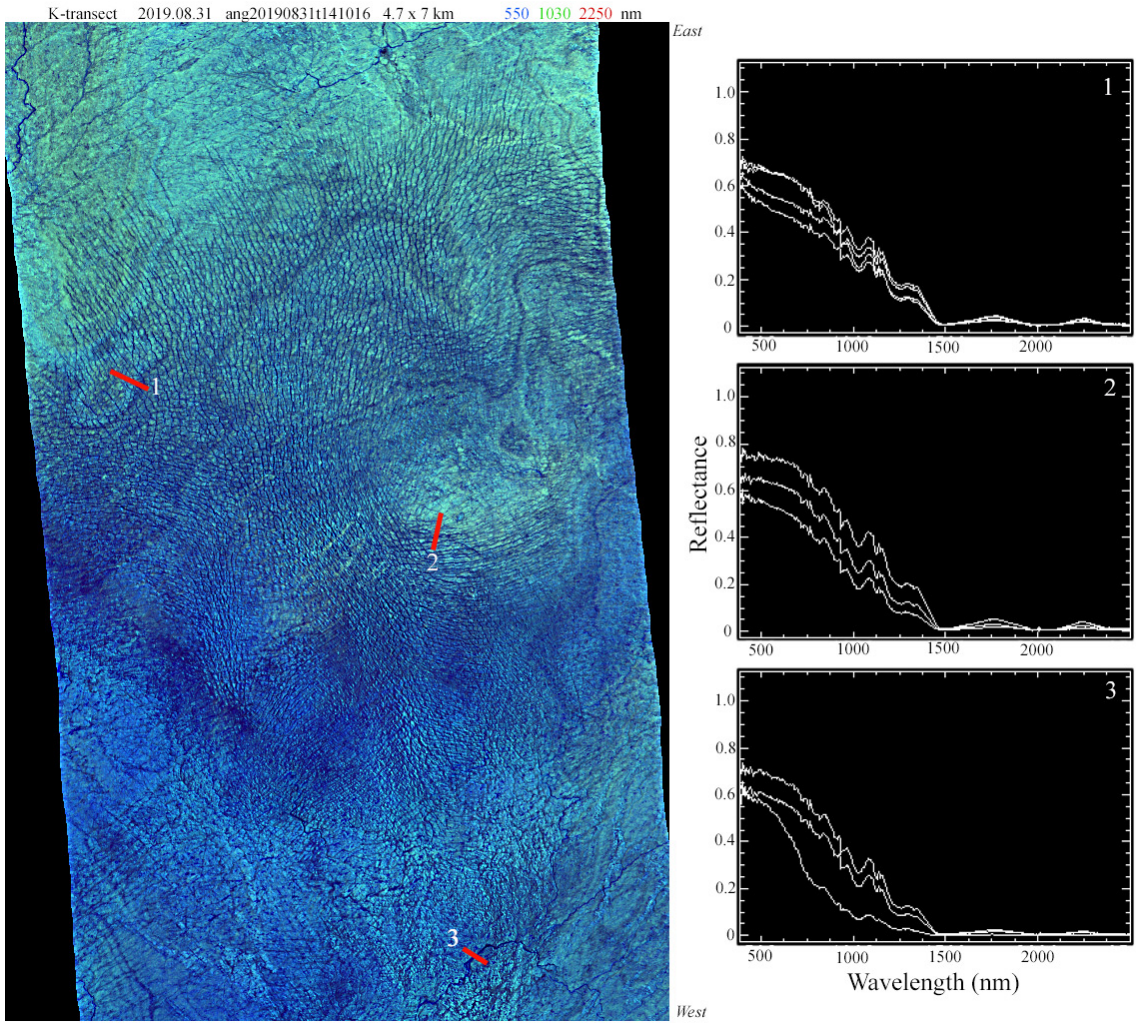

*Figure 9 Full resolution AVIRIS-NG swath and spectra from the K-transect transition zone. Variations in reflectance and illumination geometry reveal crevasse texture and lower albedo strata within the ice. Example spectra from adjacent, or near-adjacent, pixels show meter to decameter scale variations in firn and ice reflectance that would contribute to the spectral mixing occurring within decameter satellite imagery. Ground sample distance is 7.2 m.*

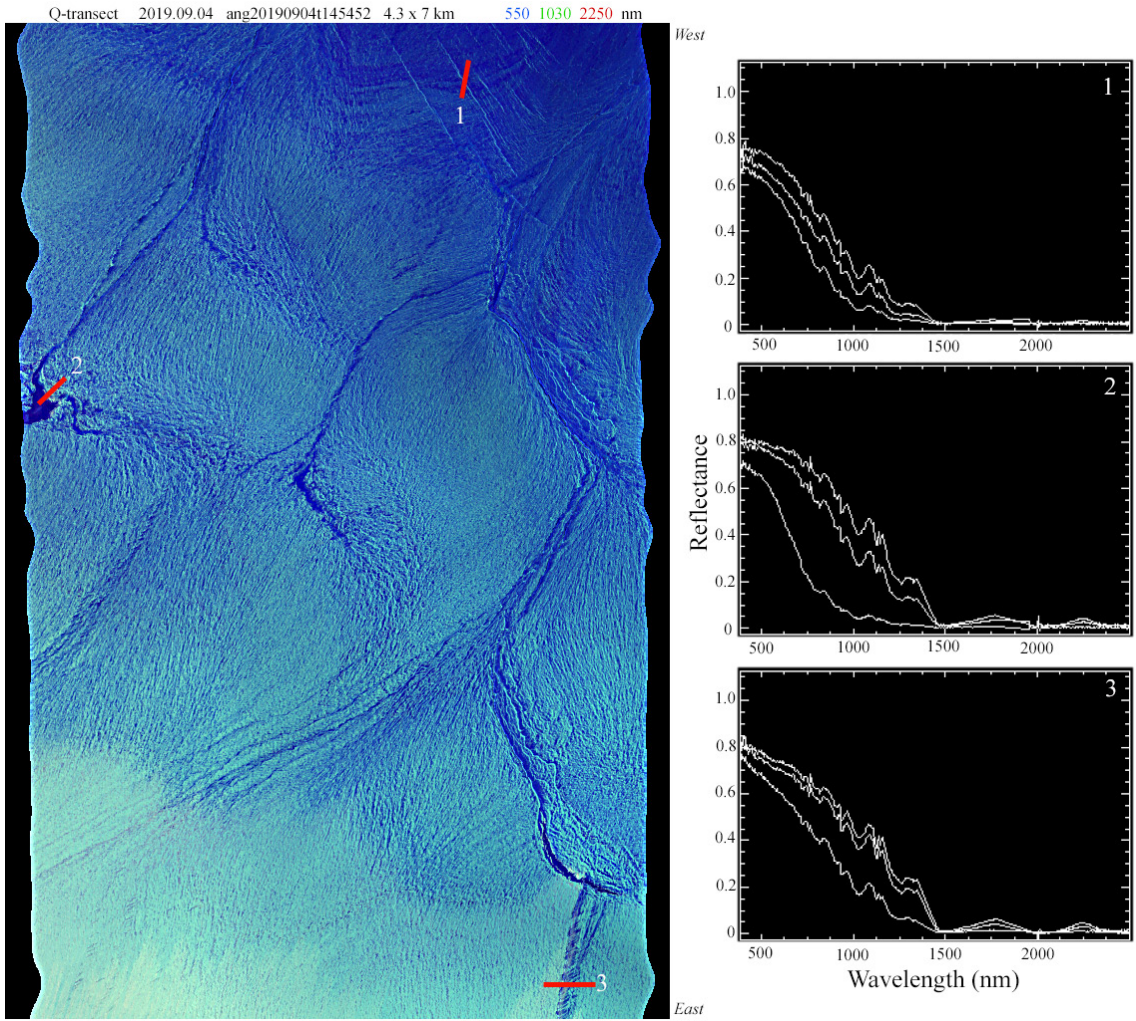

*Figure 10 Full resolution AVIRIS-NG swath and spectra from the Q-transect transition zone. Variations in reflectance and illumination geometry reveal crevasse texture and surface slope and aspect. Example spectra from adjacent, or near-adjacent, pixels show meter to decameter scale variations in ice and firn reflectance that would contribute to spectral mixing occurring in decameter resolution imagery. Ground sample distance is 6.4 m.*

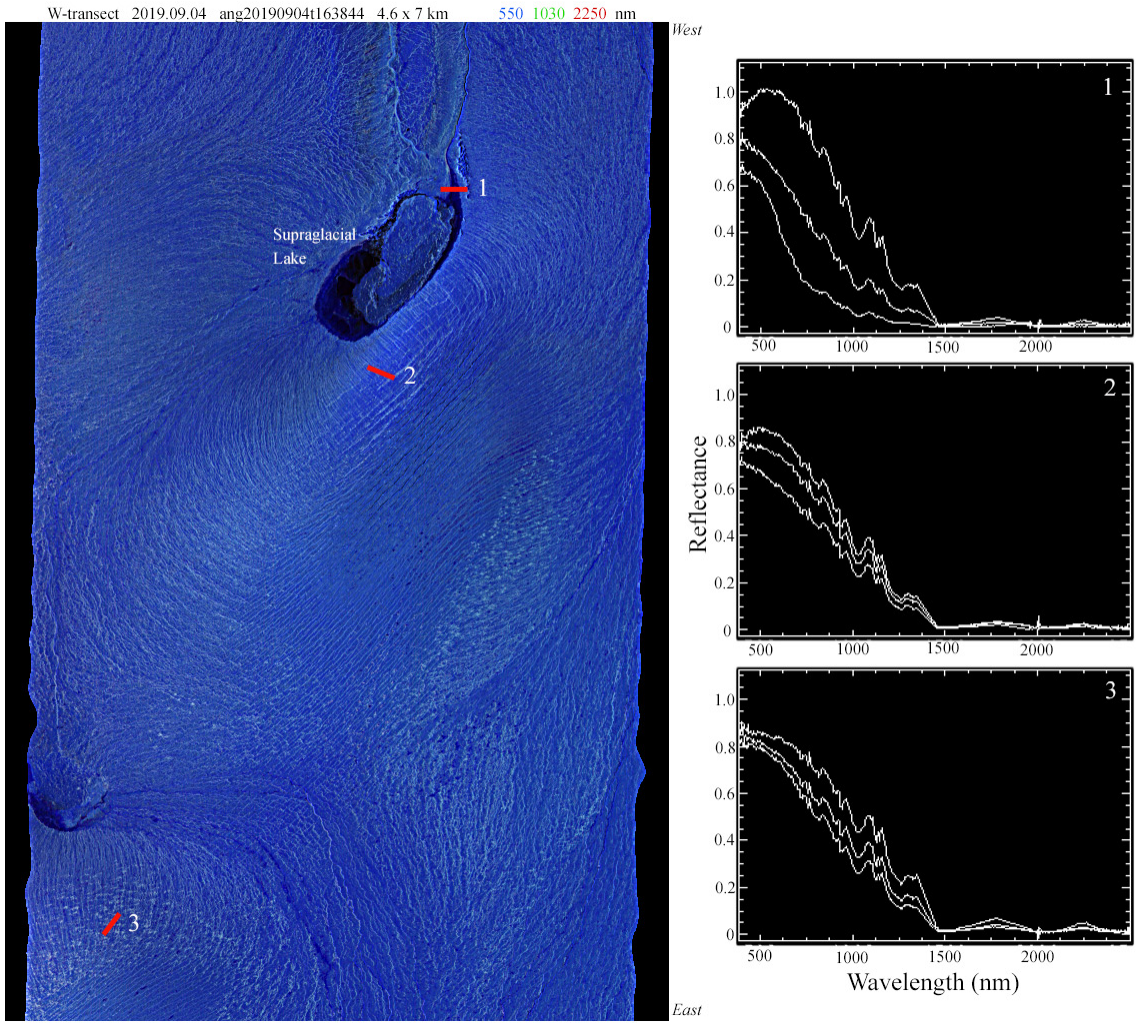

*Figure 11 Full resolution AVIRIS-NG swath and spectra from W-transect. Variations in reflectance and illumination geometry reveal crevasse texture. Example spectra from adjacent, or near-adjacent, pixels show meter to decameter scale variations in firn and ice reflectance that would contribute to the spectral mixing occurring within decameter satellite imagery. Ground sample distance is 6.8 m.*

**Discussion**

*Topology of Cryospheric Spectral Feature Spaces*

The joint characterization of the cryospheric spectral feature space combines the continuous "global" topology of the PC feature space with the clustered "local" structure of distinct manifolds captured in the t-SNE feature space. The complementarity of the global and local structure provides a basis for a generalized, physically interpretable frame of reference for cryospheric reflectance spectra from a diversity of environments while preserving distinctions among more subtle spectral characteristics resolved by the AVIRIS-NG hyperspectra.

The global feature space topology is a direct result of spectral properties of the snow-firn-ice continuum and the amplitude modulation that results from both illumination (e.g. flux density and shadow) and spectral mixing with absorptive constituents like dust and liquid water. As such, the global topology given by the PC feature space is consistent with that of the broadband Sentinel 2 feature space compiled from 50 spectrally diverse cryosphere sites by *(Small and Das 2018)*. This is to be expected because the covariance structures of both PC feature spaces are controlled by the broadband shapes and amplitudes of the snow and ice endmembers, as shown clearly by diagnostic features of the three spectral EOFs in Figure 2. Eigenvalues of the covariance matrices of both the Sentinel 2 and AVIRIS-NG composite feature spaces attribute ~99% of total variance to the two low order dimensions, making both feature spaces effectively 2D with respect to global structure. The global structure of the Sentinel 2 and AVIRIS-NG feature spaces appears to be controlled by both the grain size distribution on the snow-firn continuum, the fracture and bubble content in the glacier ice, and the illumination and spectral mixing effects modulating the spectral amplitude of the snow-firn-ice continuum. On the ice continuum, spectral amplitude and curvature are also modulated by the gradient between white and blue ice.

The local manifold structure of the cryospheric feature space rendered by t-SNE is a result of fine scale spectral characteristics like spectral curvature and narrowband absorptions (Fig. 6) that contribute little to the global covariance structure captured by the PC feature space. The fact that the clusters identified by t-SNE correspond to individual AVIRIS-NG lines (Fig. 4) and view geometry differences within individual swaths (Fig. 7) suggests a physical basis for the clustering of these manifolds within the feature space.

*PC(t-SNE) - Why it works*

Comparison of several realizations of 2D t-SNE for the composite feature space shows a consistency in the number and size distribution of clusters – despite varying shapes and positions within the 2D t-SNE space from one realization to another (Fig. 3). To establish the consistency of the spectra contained in different clusters of different realizations, we run t-SNE for an increasing number of realizations, then compute the Principal Components of suites of 2D t-SNE realizations. We hypothesize that clusters containing the same subsets of spectra in different t-SNE realizations should result in randomly located but consistently segregated t-SNE X and Y coordinates in 2D t-SNE feature spaces across multiple realizations. By treating multiple 2D t-SNE realizations as elements in each pixel's t-SNE feature vector, spectra that consistently occupy similar X and Y coordinates in 2D t-SNE feature spaces from one realization to another must also have similar t-SNE feature vectors when multiple realizations are aggregated. If the consistency of the spectra in each cluster exceeds the variability in cluster shape and position over several realizations, the feature vectors of 2D t-SNE X and Y coordinates should themselves cluster together in the larger feature space of multiple t-SNE realizations. The succession of PC(t-SNE) feature spaces shown in Figure 3 indicates that this is indeed the case as the low order dimensions of the PC(t-SNE) feature spaces become increasingly clustered as larger numbers of t-SNE realizations are included in the feature space. Back propagation of labeled clusters from the low order dimensions of PC(t-SNE) space to the individual realizations' 2D t-SNE spaces confirms the consistency of the clusters across multiple 2D t-SNE realizations while the geographic contiguity of the snow clusters reveals similarity of spectral properties. The variance partition of the PC(t-SNE) spaces suggest that 16 realizations is sufficient to achieve convergence for this AVIRIS-NG composite feature space (see Appendix for details). However, the PC(t-SNE) space used for cluster labeling and back propagation uses 30 t-SNE realizations to facilitate cluster distinction.

The PC+PC(t-SNE) feature space combines the physical interpretability of the global topology of the PC space with the local manifold structure of the PC(t-SNE) space. This joint characterization reveals distinct snow and ice continua within the space. While the PC space does show a distinct snow cluster for each transect line, these clusters are embedded within the continuous feature space without clear boundaries. With the exception of the Himalaya snow line, which maintains a distinct mixing continuum between specular reflections on sun-facing slopes, weakly illuminated snow on sun-backing slopes and exposed substrate, the upper triangular mixing space of the Greenland transects and sea ice lines forms a single continuous mixing space spanned by ice, snow and dark features (shadow and water). In contrast, the strong clustering of the PC(t-SNE) feature space clearly distinguishes multiple geographically localized continua for each line. Whereas the snow-firn continua are all geographically distinct with a different continuum for each AVIRIS-NG line, the firn-ice continua are geographically comingled across all three transect lines and the eastern sea ice line. This suggests that the spectral properties of snow-firn gradients may be more distinct than the spectral properties of firn-ice gradients. This is consistent with the structural deformation, crevassing, melting and ablation effects that are seen on the full resolution AVIRIS-NG swaths shown in Figures 8-11. In contrast to the relatively homogeneous snow cover of the accumulation zones, the fine scale deformation structure of the ices in the ablation zones juxtaposes different types of ice reflectance and illumination geometries in close proximity to each other.

*Limitations & Caveats*

While the joint characterization of global feature space topology and local manifold embeddings within the space provides insight into the properties of cryospheric spectra of different compositions, this approach does have some limitations that must be acknowledged. Foremost is the inherent non-uniqueness of data-dependent DR tools. Both the PC transform and the t-SNE manifold learning algorithms are completely data dependent, implying that changes in the input will generally result in some differences in the topology of the resulting feature space. Previous analyses of global spectral feature spaces have attempted to mitigate this data dependence by using large collections of spectrally diverse environments to approximate the actual global feature space (e.g. *(Small 2004b; Small and Milesi 2013), (Sousa and Small 2019)* and demonstrating convergence of topology, endmembers and variance partition. Clearly, the small number of AVIRIS-NG lines used in this analysis does not represent the full global diversity of cryospheric spectral properties, despite the topological similarities to the broadband feature space of the much larger and more diverse collection of Sentinel 2 spectra used by *(Small and Das 2018)*. However, upcoming missions like SBG, CHIME, and EnMap will provide global hyperspectral coverage so as to make such geographically extensive collections of cryospheric spectra feasible.

While the principal component transform does represent a type of endmember in the space of DR approaches due to its variance maximization, linearity, and orthogonality properties, the transformation is based on multiple assumptions. The primary assumption is that variance corresponds to information and that correlation implies redundancy. This may be a reasonable assumption for global topology of a feature space, but it clearly does not facilitate identification of many local scale manifolds embedded within the cyrospheric spectral feature space. Hence our interest in combining DR approaches in this joint characterization. Also, the L2 minimization upon which the PC transform is based is not the only norm possible for variance maximization. Other variations of the PC transform (e.g. Robust PCA, Kernel PCA) would be expected to produce different projections with some types of data. Even the choice of covariability metric (covariance versus correlation) can sometimes produce differing topologies – although such differences were insignificant in this analysis.

The popularity of t-SNE among non-linear DR approaches appears to derive primarily from the superiority of its clustering results, despite some idiosyncrasies that should be recognized (Wattenberg, Viégas, and Johnson 2016). The use of PC(t-SNE) compensates somewhat for the stochastic element of t-SNE, but the degree of clustering produced by t-SNE can also depend on the value of perplexity parameter chosen. While this can be addressed with a sensitivity analysis, the implication is that the clusters produced by individual t-SNE realizations are not necessarily unique – even though they are geographically consistent in the composite used in this analysis. The results presented in this analysis are based on the use of a single perplexity value of 30 (the default setting in the scikit-learn implementation), although we do provide a sensitivity analysis over a range of perplexities in the appendix below.

### *Future Applications*

The primary motivation for this analysis is to demonstrate the potential utility of joint characterization of hyperspectral feature spaces. Global analyses of the hyperspectral feature space will soon allow for identification of standardized spectral endmembers for use in linear (and non-linear) mixture models. However, the topology of the global space is dominated by the shape of the spectral continuum defining broad classes of materials. Supplementing this characterization of global feature space topology with a complementary characterization of manifolds embedded within this global space offers the potential to derive considerably greater benefit from the ability of hyperspectral imaging sensors to accurately resolve the narrow band absorptions that decorate the global space with complementary spectroscopic detail.

Until recently, hyperspectral data have been relatively sparse in both space and time. Airborne assets like AVIRIS *(Green et al., 1998)* have provided decades of critical instrument development and regional coverage, and programs like Hyperion *(Pearlman et al., 2003)* and HICO *(Corson et al., 2008)* showed important proof-of-concept for spaceborne operations, but none provided wall-to-wall global acquisition. Fortunately, this situation is destined to change in the coming years. A new generation of satellite missions are at or nearing operational status. Hyperspectral imaging is a core component of NASA's recently announced Earth System Observatory *(Margetta, 2021)*, with plans for a launch of a hyperspectral mission with global coverage and decameter spatial resolution in the mid to late 2020s. Complementary measurements from other space agencies are already being collected and/or also expected to be launched in the near future (e.g., DESIS, *Krutz et al., 2019*; PRISMA, *Candela et al., 2016*; HISUI, *Matsunaga et al., 2016*).

Global decameter hyperspectral imagery time series are thus on the horizon. Two obvious strengths of such a dataset will be systematic repeat observation of the entire Earth surface, and the establishment of an archive capable of supporting retrospective analysis. But due to the nature of many studies to date, current hyperspectral retrieval algorithms are largely designed to produce highly accurate estimates within – and potentially limited in ability to extrapolate beyond – the spatial and temporal extent of a given study. Similarly, many algorithms currently applied to global multispectral datasets are insufficient to leverage the greater dimensionality of hyperspectral data, a limitation especially notable for nonlinear optical phenomena such as those which are commonplace in cryospheric landscapes. A niche has therefore opened for a new generation of analysis tools that are capable of both high dimensionality and global generality. It is our hope that joint characterization may work towards filling this niche.

**Appendix - t-SNE hyperparameters & sensitivity analysis**

*Convergence of PC(t-SNE)*

The variance partition given by the eigenvalues of the covariance matrix provide a metric by which to assess convergence for increasing numbers of t-SNE realizations. Figure S1 shows the expected increase in curvature of the eigenvalue distribution reflecting increasing redundancy for additional realizations. As the low order dimensions account for most variance, convergence of the variance partition for the low order dimensions implies convergence of the structure of the resulting feature space of t-SNE clusters. It also implies consistency of the cluster memberships as the redundancy of additional realizations would not be expected to increase if cluster membership of spectra were random or variable from one realization to another. As shown by the inset in Figure S1, the 1, 2 and 3 dimensional spaces of the low order PC(t-SNE) dimensions all converge to effectively constant variance by 16 realizations, suggesting little benefit to additional realizations. Nonetheless, the PC(t-SNE) of the composite feature space used in this analysis is based on 30 realizations because added redundancy does cause the resulting clusters in the PC(t-SNE) feature space to coalesce and separate further making distinction and labeling easier. The rate of convergence would be expected to depend on the topology of the space and number of manifolds t-SNE resolves, so the convergence of the cryospheric composite space should not be assumed representative of other types of spectra.

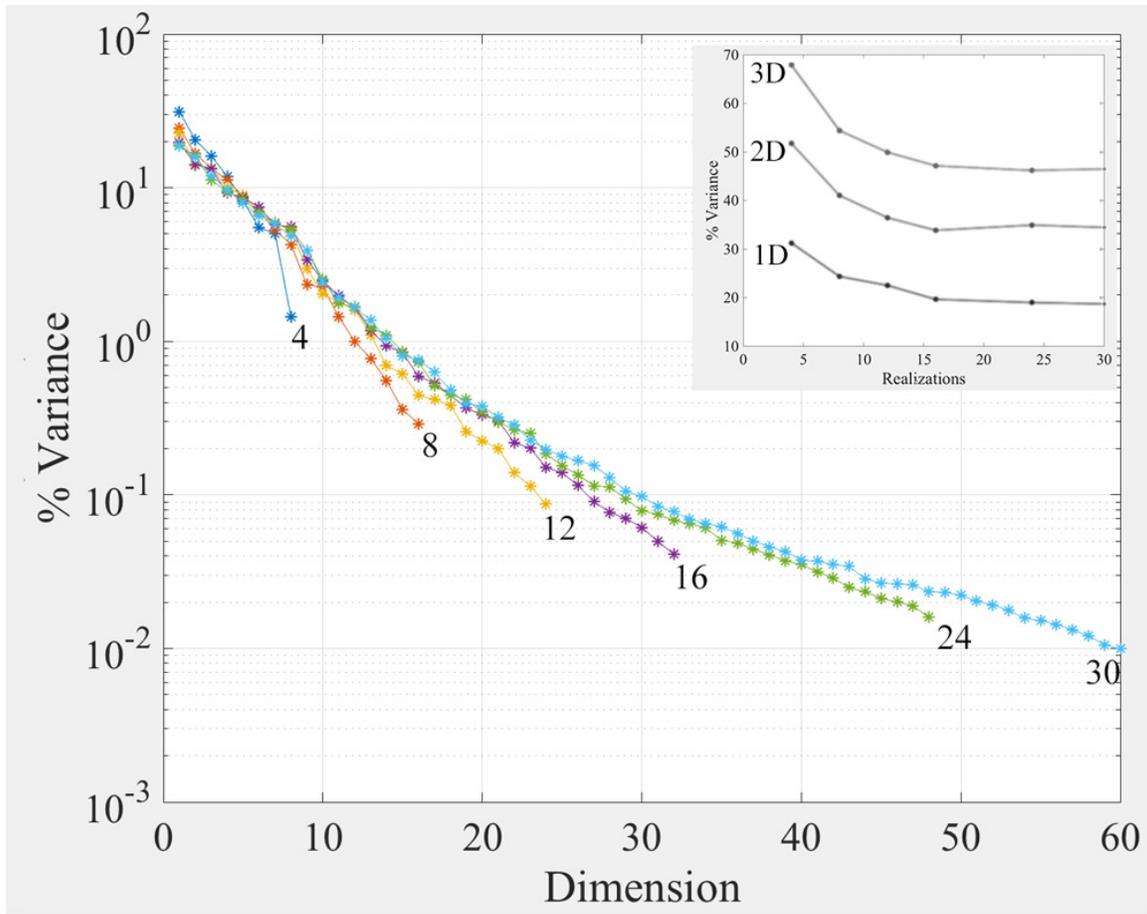

*Figure S1 Eigenvalue distributions of PC(t-SNE) for increasing 2D t-SNE realizations. Beyond 16 realizations increasing curvature exceeds decreasing slope suggesting convergence. By 30 realizations, more than half of the PC dimensions account for < 0.1% variance each. The variance accumulation for 1D, 2D and 3D projections of the low order PCs (inset) all converge by 16 realizations, suggesting little additional benefit for additional computational cost.*

*Sensitivity to Perplexity*

The perplexity parameter is generally considered to be the strongest determinant of t-SNE performance (Wattenberg, Viégas, and Johnson 2016). Increasing t-SNE perplexity generally decreases the number of clusters and increases their separability in 2D t-SNE space as larger perplexity values correspond to larger local neighborhoods over which manifolds are defined within the feature space. Incremental increases in perplexity across multiple t-SNE realizations illustrate the variation of local feature space topology as a function of perplexity, as shown by Figure S2. For perplexity > 20, the number of clusters stabilizes and separation increases. Therefore, consistency of cluster membership among t-SNE realizations with different perplexity values can be interpreted as an indication of stability and repeatability of the t-SNE feature space. Consistency of cluster membership can be demonstrated (or not) by back propagation of cluster labels from PC(t-SNE) to individual realizations of different perplexity. The 7 most distinct clusters (including dis-continua) are labeled in the three low order dimensions of PC(t-SNE) derived from all 60 (2D x 3 realizations x 10 perplexities) dimensions of 30 2D t-SNE realizations.  The results in Figure S2 are clear.  The consistency of the t-SNE cluster membership from the PC(t-SNE) feature space to the individual t-SNE realizations indicates that the cryospheric feature space of the AVIRIS-NG composite is not particularly sensitive to the perplexity parameter, although the number of clusters does not converge below a perplexity of ~20.  As with convergence of PC(t-SNE) dimensionality, the rate of convergence of perplexity should be expected to depend on the topology of the feature space being rendered.

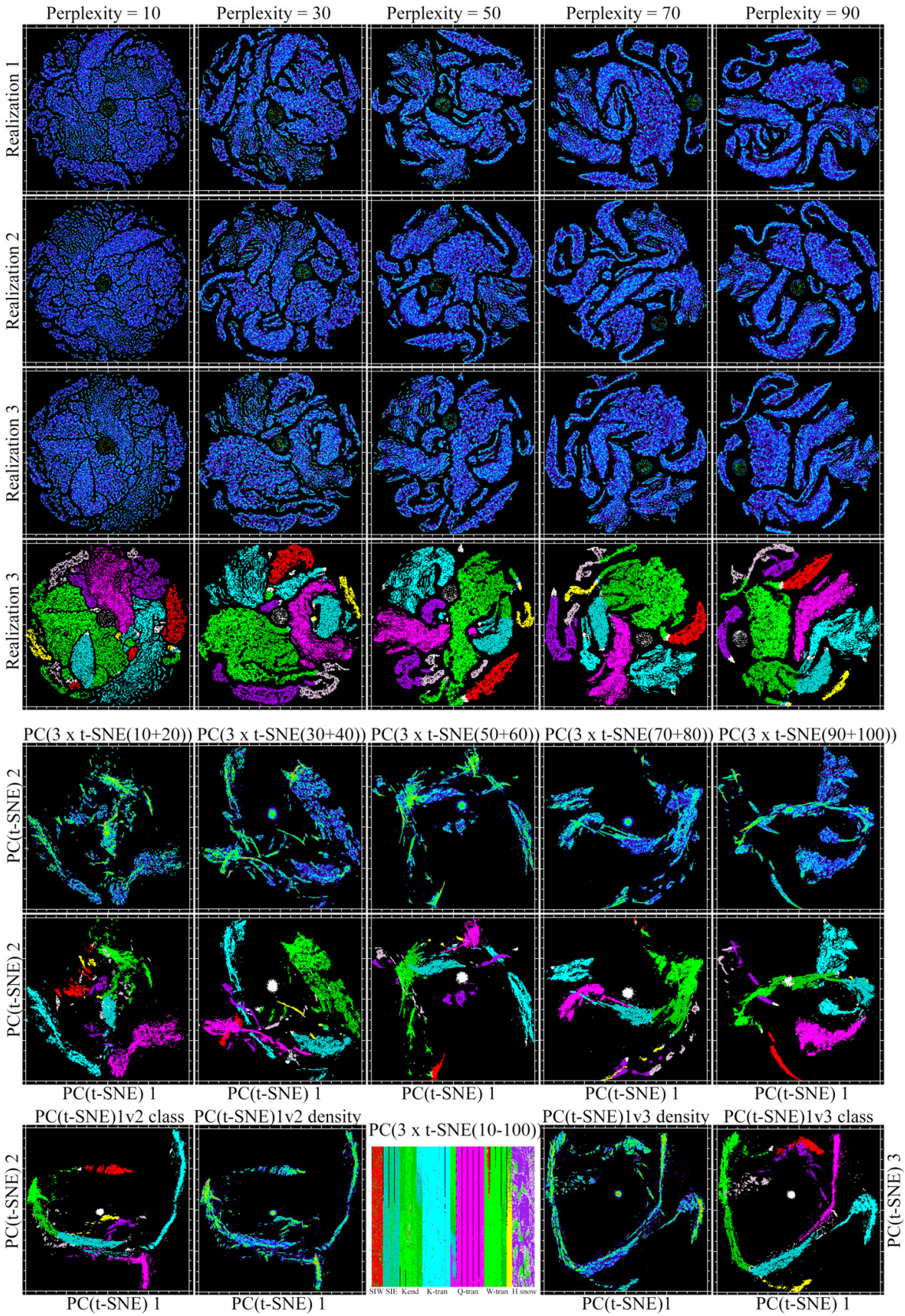

*Figure S2  Spectral feature space evolution for varying t-SNE perplexity values. Individual realizations of t-SNE with the same perplexity are distinct but share clear similarities in number and sometimes shape of clusters.  As perplexity increases, cluster separation increases and numerous smaller clusters aggregate into fewer larger clusters (top panels, L to R).  PC(t-SNE) for suites of 6 t-SNE shows a similar progression with increasing perplexity (middle panels L to R).  PC(t-SNE) for a single suite of 30 realizations spanning 10 perplexities shows similar continua and dis-continua topology (bottom panels) to the perplexity = 30 suite used in text.  Contiguous clusters (solid color labeled) in PC(t-SNE) 1v2 and 1v3 spaces generally correspond to individual AVIRIS-NG lines.  The notable exceptions are the ice clusters (greens) in which W-transect and SeaIceW comingle with western ends of both K and Q-transects.  The contiguity of these clearly distinct clusters in 30x PC(t-SNE) space can also be seen by the contiguity of label colors in the 6x PC(t-SNE) spaces and the individual t-SNE realizations labeled above.*

**Appendix - Computational specifics**

All t-SNE computations were performed on Lenovo t460s hardware running Linux (Ubuntu 18.04.6 LTS) with Intel i7-6600U CPU @ 2.60Ghz x 4, 24 GB RAM, and Intel HD Graphics 520 GPU. Computations used scikit-learn v0.24.0 in a Python 3.7 environment with the call:

> *Y = sklearn.manifold.TSNE(n_components = 2, perplexity = p).fit_transform(X)*
>
> where X is the (n x l) data matrix with l wavelengths and n pixels, and Y is the (n x 2) t-SNE output matrix.

Perplexity variation experiments on the subsampled data cube (n = 17164 pixels with 373 channels) found runtime sensitivity to perplexity value to be sublinear (runtime approximately 75 seconds for perplexity = 10;  90 seconds for perplexity = 50; 130 seconds for perplexity = 100).

Small, C., and I. Das. 2018. "The cryospheric spectral mixing space." In *American Geophysical Union Fall Meeting*. Washington DC: American Geophysical Union.
Small, C., and C. Milesi. 2013. 'Multi-scale Standardized Spectral Mixture Models', *Remote Sensing of Environment*, 136: 442-54.
Sousa, D., and C. Small. 2018. 'Multisensor analysis of spectral dimensionality and soil diversity in the Great Central Valley of California', *Sensors*, 18: 1-17.
———. 2019. 'Globally standardized MODIS spectral mixture models', *Remote Sensing Letters*, 10: 1018-27.
———. 2021. 'Joint Characterization of Multiscale Information in High Dimensional Data', *ArXiv*, 2102.09669.
Thompson, D.R., J.W. Boardman, M.L. Eastwood, and R.O. Green. 2017. 'A large airborne survey of Earth's visible-infrared spectral dimensionality', *Optics Express*, 25: 9186-95.
Warren, S.G. 2013. 'Can black carbon in snow be detected by remote sensing?', *Journal of Geophysical Research*, 118: 779-86.
———. 2019. 'Optical properties of ice and snow', *Philosophical Transactions of the Royal Society A*, 377.
Warren, S.G., and W.J. Wiscombe. 1980. 'A model for the spectral albedo of snow II: Snow containing atmospheric aerosols', *Journal of the Atmospheric Sciences*, 37.
Wattenberg, M., F. Viégas, and I. Johnson. 2016. 'How to Use t-SNE Effectively', *Distill*.
Wiscombe, W.J., and S.G. Warren. 1980. 'A model for the spectral albedo of snow I: Pure Snow', *Journal of the Atmospheric Sciences*, 37: 2712-45.
30